\RequirePackage{ifpdf}
\ifpdf 
\documentclass[pdftex]{sigma}
\else
\documentclass{sigma}
\fi

\begin{document}

\allowdisplaybreaks

\renewcommand{\thefootnote}{$\star$}

\renewcommand{\PaperNumber}{057}

\FirstPageHeading

\ShortArticleName{Symmetry Operators for Dirac's Equation}

\ArticleName{Symmetry Operators and Separation of Variables for Dirac's Equation on Two-Dimensional Spin Manifolds\footnote{This paper is a
contribution to the Special Issue ``Symmetry, Separation, Super-integrability and Special Functions~(S$^4$)''. The
full collection is available at
\href{http://www.emis.de/journals/SIGMA/S4.html}{http://www.emis.de/journals/SIGMA/S4.html}}}

\AuthorNameForHeading{A. Carignano, L. Fatibene, R.G. McLenaghan and G. Rastelli}

\Author{Alberto CARIGNANO~$^{\dag^1}$, Lorenzo FATIBENE~$^{\dag^2}$, Raymond G. McLENAGHAN~$^{\dag^3}$\\ and Giovanni RASTELLI~$^{\dag^4}$}

\Address{$^{\dag^1}$~Department of Engineering, University of Cambridge, United Kingdom}
\EmailDD{\href{mailto:ac737@cam.ac.uk}{ac737@cam.ac.uk}}

\Address{$^{\dag^2}$~Dipartimento di Matematica, Universit\`{a} di Torino, Italy}
\EmailDD{\href{mailto:lorenzo.fatibene@unito.it}{lorenzo.fatibene@unito.it}}

\Address{$^{\dag^3}$~Department of Applied Mathematics, University of Waterloo, Ontario, Canada}
\EmailDD{\href{mailto:rgmclena@uwaterloo.ca}{rgmclena@uwaterloo.ca}}

\Address{$^{\dag^4}$~Formerly at Dipartimento di Matematica, Universit\`{a} di Torino, Italy}
\EmailDD{\href{mailto:giorast.giorast@alice.it}{giorast.giorast@alice.it}}

\ArticleDates{Received February 01, 2011, in f\/inal form June 02, 2011;  Published online June 15, 2011}

\Abstract{A signature independent formalism is created and utilized to determine the ge\-ne\-ral
second-order symmetry operators for Dirac's equation on two-dimensional Lorentzian spin manifolds.
The formalism is used to characterize the orthonormal frames and metrics that permit
the solution of Dirac's equation by separation of variables in the case where a~second-order
symmetry operator underlies the separation.  Separation of variables in complex variables on
two-dimensional Minkowski space is also considered.}


\Classification{70S10; 81Q80}

\begin{flushright}
\begin{minipage}{96mm}
\it This paper is dedicated to Professor Willard Miller,~Jr. on the occasion of his retirement from the School of Mathe\-matics at the University
of Minnesota.
\end{minipage}
\end{flushright}

\section{Introduction}

\looseness=-1
This paper is a contribution to the study of the separability theory for Dirac's equation to which Professor Miller has made important
contributions \cite{KMW1,KMW2,MILLER2}.
Exact solutions to Dirac's re\-la\-tivistic wave equation by means of the method of separation of variables have been studied since the equation was postulated in 1928.  Indeed, the solution for the hydrogen atom may be obtained by this method.  While there is a well developed theory of separation of variables for the Hamilton--Jacobi equation, and the
Schr\"{o}dinger equation based on the existence of valence two characteristic Killing tensors which def\/ine
respectively quadratic f\/irst integrals and second-order symmetry operators for these equations (see \cite{MILLER1,KALNINS,B97,FFFG})
an analogous theory for the Dirac equation is still in its early stages.  The complications arise from the fact that one is dealing with a~system of f\/irst-order partial dif\/ferential equations whose derivation from the invariant Dirac equation depends not only on the choice of coordinate system but also on the choice of an orthonormal moving frame and  representation for the Dirac matrices with respect to which the components of the unknown spinor are def\/ined.  Further complications arise if the background space-time is assumed to be curved.  Much of the progress in the theory  has been stimulated by developments in Einstein's general theory of relativity where one studies
f\/irst quantized relativistic electrons on curved background space-times of physical interest such as the Schwarzschild and Kerr black hole space-times.  This work required the preliminary development of a theory of spinors  on  general pseudo-Riemmanian manifolds (see \cite{FF,CFF,Jad, Spi}).  The solution of the Dirac equation in the Reissner--Nordstrom solution was apparently f\/irst obtained by Brill and Wheeler in 1957~\cite{BW} who separated the equations for the spinor components in standard orthogonal Schwarzschild coordinates with respect to a moving frame adapted to the coordinate curves.  A comparable separable solution in the Kerr solution was found by Chandrasekhar in 1976~\cite{CHAN} by use of an ingenious separation ansatz involving Boyer--Lindquist coordinates and the Kinnersley tetrad.  The separability property was characterized invariantly by Carter and
McLenaghan~\cite{CM} in terms of a f\/irst-order dif\/ferential operator constructed from the valence two Yano--Killing tensor that exists in the Kerr solution, that commutes with the Dirac operator and that admits the separable solutions as eigenfunctions with the separation constant as eigenvalue.   Study of this example led Miller~\cite{MILLER2} to propose the theory of a factorizable system for f\/irst-order systems of Dirac type in the context of which the separability property may be characterized by the existence of a certain system of commuting f\/irst-order symmetry operators.   While this theory includes the Dirac equation on the Kerr solution and its generalizations~\cite{FK} it is apparently not complete since as is shown by Fels and Kamran~\cite{FK} there exist systems of the Dirac type whose separability is characterized by second-order symmetry operators.  The work begun by these authors has been continued by Smith~\cite{Sm},
Fatibene, McLenaghan and Smith~\cite{FMS}, McLenaghan and Rastelli~\cite{MR}, and Fatibene, McLenaghan, Rastelli and Smith~\cite{FMRS} who studied the problem in the simplest
possible setting namely on two-dimensional Riemmanian spin manifolds.
The motivation for working in
the lowest permitted dimension is that it is possible to examine in detail the dif\/ferent possible scenarios that arise from the separation ansatz and the imposition of the separation paradigm that the separation be characterized by a symmetry operator admitting the separable solutions as eigenfunctions.
The insight obtained from this approach may help suggest an approach to take for the construction of a general separability theory for Dirac type equations.  Indeed in~\cite{MR} systems of two f\/irst-order linear partial equations of Dirac type which admit multiplicative separation of variables in some arbitrary coordinate system and whose separation constants are associated with commuting dif\/ferential operators are exhaustively characterized.  The requirement
that the original system arises from the Dirac equation on some two-dimensional Riemannian spin manifold allows the local
characterization of the orthonormal frames and metrics admitting separation of variables for the equation and the determination of the symmetry operators associated to the separation.  The paper~\cite{FMRS} takes this research in a dif\/ferent but closely related direction.  Following
earlier work of McLenaghan and Spindel~\cite{MSp} and Kamran and McLenaghan~\cite{KML} where the f\/irst-order
symmetry operators of the Dirac equation where computed on four-dimensional Lorentzian spin manifolds and McLenaghan,
Smith and Walker~\cite{MSW} where the second-order operators were determined in terms of a two-component spinor formalism,
the most general second-order linear dif\/ferential operator which commutes with the Dirac operator on a general
two-dimensional Riemannian spin manifold is obtained.  Further it is shown that the operator is characterized in terms of Killing vectors and valence two Killing
tensors def\/ined on the background manifold.  The derivation is manifestly covariant: the calculations are done in a general orthonormal frame without the choice of a particular
set of Dirac matrices.

The purpose of the present paper is to extend the results of~\cite{MR} and~\cite{FMRS} described above to the case of two-dimensional Lorentzian spin manifolds.
One of the main achievements of the paper is the creation of a formalism that permits the {\em  simultaneous treatment of both} the Riemannian and Lorentzian cases.
Further, we extend the results of \cite{DR}, where Hamilton--Jacobi separability separability is studied in complex variables on two-dimensional Minkowski space, to the Dirac equation.

The paper is organized as follows. In Section~\ref{section2} we summarize the basic properties of two-dimensional spin manifolds required for the subsequent calculations.
Section~\ref{section3} is devoted to the derivation of the form of the general second-order linear dif\/ferential operator which commutes with the Dirac operator.
We show that this operator is characterized by a valence two Killing tensor f\/ield, two Killing vector f\/ields and two scalar f\/ields def\/ined on the
background spin-manifold.
In Section~\ref{section4} we develop a formalism based on \cite{MR} which enables us to study simultaneously separation of variables for the Dirac equation on {\em both}
Riemannian and Lorentzian spin manifolds.  All possible cases where separation occurs are determined.
In Section~\ref{section5} we establish a link between the second-order symmetry operators obtained formally in Section~\ref{section3} and the second-order symmetry operator underlying the
separation of variables scheme considered in  Section~\ref{section4}.
In Section~\ref{section6} separation of Dirac's equation in complex variables is considered on two-dimensional Minkowski space.
Section~\ref{section7} contains the appendix.  The notation and conventions of this paper  are consistent with~\cite{FMRS}.

\section{Framework}\label{section2}

Let $M$ be a connected, paracompact, two-dimensional spin manifold.
Let us consider both the Euclidean $\eta=(2,0)$  and the Lorentzian $\eta=(1,1)$ signatures. With an abuse of notation, let~$\eta$ also denote the canonical form induced by the signature and the determinant of such quadratic form, namely one has $\eta=\pm1$. We will keep this sign as an undetermined parameter in this paper, since we want to consider both cases at once.

We know that a representation of the Clif\/ford algebra is induced by a set of Dirac matrices~$\gamma_a$ such that they satisfy the Dirac condition
\begin{gather*} \gamma_a\gamma_b+\gamma_b\gamma_a = 2 \eta_{ab}I. 
\end{gather*}
The generators of the even Clif\/ford algebra are $I$, $\gamma_a$ and $\gamma:=\gamma_1\gamma_2$.
Therefore the most general element of the ${\rm Spin}(\eta$) group is
\[ S=a I+b\gamma \]
with $a^2+\eta b^2=1$.

From the theory, we know that it is possible to def\/ine a covering map between ${\rm Spin}(\eta)$ and~${\rm SO}(\eta)$. Let $l^a_b$ be a generic element of ${\rm SO}(\eta)$. Then
\[ \eta_{ab}=l^c_a\eta_{cd}l^d_b  \]
and the covering map in matrix form is
\[ l(S)=   \begin{pmatrix}
 a^2-\eta b^2 & 2\eta ab  \\
 -2\eta ab & a^2-\eta b^2
 \end{pmatrix}.   \]
Let $P\rightarrow M$ be a suitable ${\rm  Spin}(\eta$) principal bundle, such that it allows global maps $\Lambda:P \rightarrow L(M)$ of the spin bundle into the general frame bundle of $M$. The local expression of such maps is given by spin frames $e^\mu_a$.

A spin frame induces a metric of signature $\eta$ and the corresponding spin connection:
\[ g_{\mu\nu}=e_{\mu}^a\eta_{ab}e_{\nu}^b,
 \qquad \Gamma^{ab}_{\mu}=e^a_{\alpha}\big(\Gamma^{\alpha}_{\beta\mu}e^{b\beta}+\partial_{\mu}e^{\alpha}_c\eta^{cb}\big) , \]
where $\Gamma^{\alpha}_{\beta\mu}$ denotes the Levi-Civita connection of the induced metric $g_{\mu\nu}$, which in turn induces the covariant derivative as $\nabla_{\mu}=e^a_{\mu}\nabla_a$.

It follows from our setting that the inner product $\eta_{ab}$ will raise and lower Latin indices, while the Greek ones are raised and lowered by the induced metric $g_{\mu\nu}$.

Finally, the covariant derivative of spinors is def\/ined as
\begin{gather}\nabla_{\mu}\psi:=\partial_{\mu}\psi+\Gamma_{\mu}\psi,
 \label{der}  \end{gather}
where
\[
\Gamma_{\mu}= \tfrac{1}{8} \Gamma^{ab}_{\mu}[\gamma_a,\gamma_b]=\tfrac{1}{4} \Gamma^{ab}_{\mu}\epsilon_{ab}\gamma .
  \]
The covariant derivative (\ref{der}) is invariant under spin transformations in view of the following property
\begin{gather*}
 S\partial_{\nu}S^{-1}=\tfrac{1}{4} l^b_a\partial_{\nu}\bar{l}_c^a\epsilon_b{}_\cdot^c\gamma.
 \end{gather*}
A spin transformation is an automorphism of the spin bundle $P$. If we def\/ine the f\/iber coordinates of $P$ as $(x, S)$, the local expressions of the spin transformation are
$x'^{\mu}=x'^{\mu}(x)$, $S'=\phi(x)S$,
where $\phi:U\rightarrow {\rm Spin}(\eta)$.

As we know, it acts as a left group on spinors and spin frames ($l_a^b\equiv l_a^b(\varphi)$)
\begin{gather*}  \psi'=\phi\psi, \qquad e'^{\mu}_a=J^{\mu}_{\nu}e^{\nu}_b\bar{l}^b_a,
\end{gather*}
where $l:{\rm Spin}(\eta)\rightarrow {\rm SO}(\eta)$ and $J^{\mu}_{\nu}$ is the Jacobian of the coordinate transformation.
Therefore the spin connection transforms as
\begin{gather*} \Gamma'^{ab}_{\mu}=\bar{J}^{\nu}_{\mu}l^a_c(\psi)\big(\Gamma^{cd}_{\nu}l^b_d(\psi)+\partial_{\nu}\bar{l}^c_d(\psi)\eta^{db}\big).
\end{gather*}
Looking back to the covariant derivative, one can prove that the commutator can be related to the curvature
\begin{gather}
[\nabla_{\mu},\nabla_{\nu}]\psi=\tfrac{1}{4} \gamma\psi\epsilon_{ab}R^{ab}_{\,\,\mu\nu} .
\label{com}
\end{gather}
The advantage of working in dimension two is that the Riemann tensor $R^{\alpha\beta}_{\,\,\mu\nu}$ has only one independent component that can be written as a function of the Ricci scalar~$R$.
Hence the identity~(\ref{com}) may be rewritten as
\begin{gather}
 [\nabla_{c},\nabla_{d}]\psi=\tfrac{1}{4} \gamma\psi\epsilon_{cd}R.
 \label{com1}
 \end{gather}
Similarly the following property holds
\begin{gather} [\nabla_a,\nabla_b]\nabla_c\psi=\tfrac{R}{4} \gamma\epsilon_{ab}\nabla_{c}\psi-\tfrac{R}{2} \epsilon^d{}^{\cdot}_{c}\epsilon_{ab}\nabla_d\psi . \label{com2}
\end{gather}

\section{First- and second-order symmetry operators}\label{section3}

In our framework, the Dirac equation has the form
\begin{gather}
\mathbb{D}\psi=i\gamma^a\nabla_a\psi-m\psi=0. \label{Dirac}
\end{gather}
An operator is a symmetry operator for the Dirac equation if
\begin{gather}  [\mathbb{K},\mathbb{D}]=0 . \label{sym1}
\end{gather}
The most general operator of the second-order has the form
\begin{gather*} \mathbb{K}=\mathbb{E}^{ab}\nabla_{ab}+\mathbb{F}^a\nabla_a+\mathbb{G}I,
\end{gather*}
where $\mathbb{E}^{ab}$, $\mathbb{F}^a$, $\mathbb{G}$ are algebraic matrix coef\/f\/icients to be determined.

We aim to determine coef\/f\/icients so that condition \eqref{sym1} holds true.
Considering the Ricci's identities~\eqref{com},~\eqref{com1} and~\eqref{com2}, we expand the symmetry equation~\eqref{sym1} for the coef\/f\/icients and we obtain
\begin{gather}
\mathbb{E}^{(ab}\gamma^{c)}-\gamma^{(c}\mathbb{E}^{ab)}=0, \qquad
\mathbb{F}^{(a}\gamma^{b)}-\gamma^{(b}\mathbb{F}^{a)}=\gamma^c\nabla_c \mathbb{E}^{ab},\nonumber\\
\mathbb{G}\gamma^d-\gamma^d \mathbb{G}=\gamma^c \nabla_c \mathbb{F}^d-(\mathbb{E}^{ad}\gamma^c+\gamma^c \mathbb{E}^{ad})\gamma\tfrac{R}{4} \epsilon_{ac}+ +\tfrac{R}{3} \big(\tfrac{1}{2} \mathbb{E}^{ba}\gamma^c+\gamma^c \mathbb{E}^{ba}\big)\epsilon^d{}^{\cdot}_a\epsilon_{bc},\nonumber\\
\gamma^c \nabla_c \mathbb{G}=\tfrac{R}{8} \left(\mathbb{F}^a\gamma^c +\gamma^c \mathbb{F}^a\right)\gamma \epsilon_{ac}+\big(\tfrac{1}{2} \gamma^c \mathbb{E}^{ab}+\mathbb{E}^{ab}\gamma^c\big)\tfrac{1}{6} \gamma\epsilon_{bc}\nabla_b R.
\label{cond}
\end{gather}
To write the system (\ref{cond}) we make use of a characterization of second- and third-order covariant derivatives which is shown in the Appendix.

Let us begin with the f\/irst equation in \eqref{cond}. We know that the coef\/f\/icients of $\mathbb{E}^{ab}$ are zero-order matrix operators (i.e.\ not dif\/ferential) which can be expanded in the basis of $C(\eta)$
\begin{gather*}
 \mathbb{E}^{ab}=e^{ab}I+e^{ab}_c\gamma^c+\hat{e}^{ab}\gamma,
 \end{gather*}
where the coef\/f\/icients are point functions in $M$. Here we are using the fact that $I$, $\gamma^c$ and $\gamma$ form a basis, since they are linearly independent.
Hence, the f\/irst equation can be rewritten as
\begin{gather*} -2\hat{e}^{(ab}\epsilon^{c)\cdot}_{\,\,d}\gamma^d-2\eta e^{(ab}_d\epsilon^{dc)}\gamma=0,
 \end{gather*}
which can be solved to obtain
\begin{gather*} \hat{e}^{ab}=0, \qquad   e^{ab}_d\gamma^d=2\alpha^{(a}\gamma^{b)}, \end{gather*}
where $\alpha^a$ are the coef\/f\/icients along the frame $e_a:=e^{\mu}_a\partial_{\mu}$ of an arbitrary vector f\/ield~$\alpha$.

Applying these conditions, the operator $\mathbb{E}$ is of the form
\begin{gather*} \mathbb{E}^{ab}=e^{ab}I+2\alpha^{(a}\gamma^{b)}. 
\end{gather*}
Now we consider the second condition of~\eqref{cond}. As we did for $\mathbb{E}$, we expand $\mathbb{F}$ in the basis
\begin{gather*}
 \mathbb{F}^{a}=f^{a}I+f^a_b\gamma^b+\hat{f}^{a}\gamma, 
\end{gather*}
and substitute it back in the condition \eqref{cond}. We obtain
\begin{gather}
\nabla_c \alpha^{(a}\eta^{b)c}=0,\qquad
f^{(a}_c\epsilon^{b)c}=\nabla_c \alpha^{(a}\epsilon^{b)c},\qquad
-2\hat{f}^{(a}\epsilon^{b)\cdot}_{\,\,c}=\nabla_c e^{ab}.
\label{con2}
\end{gather}
The f\/irst equation says that $\alpha^a$ is a Killing vector of $g$ or possibly the zero vector. Through an explicit calculation, we prove that the most general tensor satisfying the second condition is~$A\delta^a_c$. Hence
\begin{gather*}
f^a_c=\nabla_c \alpha^a+A\delta^a_c.
    \end{gather*}
Finally, we substitute into the third equation of \eqref{con2} obtaining
\begin{gather}
-2\hat{f}^{(a}\epsilon^{b)\cdot}_{\,\,c}
\epsilon^c{}
_{\cdot d}=\nabla_c e^{ab}
\epsilon^c{}_{\cdot d}. \label{con3}
\end{gather}
Taking the trace we get
\begin{gather} \hat{f}^b=\tfrac{1}{3} \nabla_ce^{ab}\epsilon^{c}_{\cdot a}=\tfrac{1}{3} \nabla^ce^{ab}\epsilon_{ca},  \label{con3b}
\end{gather}
which shows that $\hat{f}^a$ is uniquely determined by $e^{ac}$. However~\eqref{con3} contains six equations, of which only one has been used. The other f\/ive equations are exploited by back substituting~\eqref{con3b} into~\eqref{con3} to obtain an equation for $e^{ac}$ alone, namely
\begin{gather*}
\nabla^c e^{ea}\epsilon_{ce}\delta^b_d+\nabla^c e^{eb}\epsilon_{ce}\delta^{a}_d=\nabla_c
e^{ab}\epsilon^{c}_{\cdot\,\,d}.
\end{gather*}  	
This is an integrability condition for $e^{ab}$ which may be written as
\begin{gather*}
\nabla^{(a}e^{bc)}=0. 
\end{gather*}
This equation shows that $e^{ab}$ is a Killing tensor of $g$.

We shall now consider the third condition of \eqref{cond}
 \begin{eqnarray*}
\mathbb{G}\gamma^d-\gamma^d \mathbb{G}=\gamma^c \nabla_c \mathbb{F}^d-(\mathbb{E}^{ad}\gamma^c+\gamma^c \mathbb{E}^{ad})\gamma\tfrac{R}{4}\epsilon_{ac}+\tfrac{R}{3}\big(\tfrac{1}{2}\mathbb{E}^{ba}\gamma^c+\gamma^c \mathbb{E}^{ba}\big)\epsilon^d{}^{\cdot}_a\epsilon_{bc}.
\end{eqnarray*}

Considering the usual expansion $\mathbb{G}=gI+g_a\gamma^a+\hat{g}\gamma$, we obtain the following system of equations
\begin{gather*}
0=\nabla_a A-\tfrac{R}{2} \alpha^aI+\tfrac{R}{2} \alpha^aI,\\
2g_b\epsilon^{ba}_{\cdot\cdot}=\big(\nabla_b A-\tfrac{R}{2} \alpha_b\big)\epsilon^{ba}_{\cdot\cdot}-\tfrac{R}{2}
\alpha_b\epsilon^{ba}_{\cdot\cdot}-\tfrac{R}{2} \alpha_b \epsilon^{ab}_{\cdot\cdot},\\
-2\eta\hat{g}\epsilon^{ab}_{\cdot \cdot}\gamma^b=\nabla_b f^a+\tfrac{\eta}{3} \epsilon_{dc}\epsilon_e{}_{\cdot}^b\nabla^e\nabla^d e^{ac}+\tfrac{R}{2} e^{ab}+\tfrac{R}{2} e^{cd}\epsilon^a{}^{\cdot}_d\epsilon_c{}_{\cdot}^b.
\end{gather*}
The f\/irst equation means that $A\in\mathbb{C}$, while the second implies $g_b=-\tfrac{R}{4} \alpha_b$. By splitting the third into its symmetric and antisymmetric parts, it results that
\begin{gather*}
\hat{g}=\tfrac{1}{4} \epsilon_{ba}\nabla^b f^a 
\end{gather*}
for the antisymmetric part, while the symmetric part can be expanded to obtain
\begin{gather*}  \nabla^{(a}(f^{b)}-\nabla_c e^{b)c})=0. 
\end{gather*}
It follows that there exists a Killing vector or a zero-vector $\zeta$ such that $\zeta^a=f^a-\nabla_c e^{ac}$ from which we obtain
\begin{gather*}
f^a=\zeta^a+\nabla_c e^{ac}. 
\end{gather*}
We summarize all the results obtained so far
\begin{gather*} \mathbb{E}^{ab}=e^{ab}I+2\alpha^{(a}\gamma^{b)},\\
 \mathbb{F}^a=(\zeta^a+\nabla_c e^{ac})I+(\gamma^c\nabla_c\alpha^a+A\gamma^a)+\tfrac{1}{3} \epsilon_{bc}\nabla^b e^{ac}\gamma,\\
 \mathbb{G}=gI-\tfrac{R}{4} \alpha_b\gamma^b+\tfrac{1}{4} \epsilon_{ba}\nabla^b\zeta^a\gamma,
\end{gather*}
where
\begin{gather*}  A\in \mathbb{C}, \qquad
 \nabla^{(d}e^{ab)}=0,\qquad
 \nabla^{(a}\alpha^{b)}=0, \qquad  \nabla^{(d}\zeta^{b)}=0.
\end{gather*}
It remains to consider the fourth equation in \eqref{cond}
\begin{gather*}
\gamma^c \nabla_c \mathbb{G}=\tfrac{R}{8} \left(\mathbb{F}^a\gamma^c +\gamma^c \mathbb{F}^a\right)\gamma \epsilon_{ac}+\big(\tfrac{1}{2} \gamma^c \mathbb{E}^{ab}+\mathbb{E}^{ab}\gamma^c\big)\tfrac{1}{6} \gamma\epsilon_{bc}\nabla_b R,
\end{gather*}
which implies the only additional condition:
\begin{gather} \nabla^a g=-\tfrac{1}{4} \nabla_b(Re^{ab}). \label{integra}
\end{gather}
This equation locally determines $g$ if and only if the right hand side is a closed 1-form. We call~\eqref{integra} the \emph{integrability condition}.

We now focus on f\/inding condition for f\/irst-order operators.

We can easily obtain the conditions by setting to zero $e^{ab}$ and $\alpha^a$. In particular \eqref{integra} is trivially satisf\/ied and $g \in \mathbb C$. We thus obtain that the most general f\/irst-order symmetry operator may be written as
\begin{gather*}
\mathbb{E}^{ab}=0,\qquad
\mathbb{F}^a=\zeta^aI+A\gamma^a,\qquad
\mathbb{G}=gI+\tfrac{1}{4} \epsilon_{ba}\nabla^b \zeta^a\gamma,
\end{gather*}
where $A, g \in\mathbb C$ and $\nabla^{(d}\zeta^{b)}=0$.

\section{Separation of variables}\label{section4}

Let us start with the Dirac condition
\begin{gather}
\gamma^a\gamma^b+\gamma^b\gamma^a = 2 \eta^{ab}I .	\label{Dir}
\end{gather}
We f\/ix the metric convention to be $\eta^{ab}={\rm diag}(1,\pm1)$.

A choice of gamma matrices valid for both signatures is
\begin{gather*}
\gamma^1=  \begin{pmatrix}
 1 & 0  \\
 0 & -1
 \end{pmatrix},\qquad
 \gamma^2=\begin{pmatrix}
 0 & -k  \\
 k & 0
\end{pmatrix}, 
\end{gather*}
where $k=i$ for Euclidean and $k=1$ for Lorentzian signature.

Using the gamma matrices and (\ref{der}), we can rewrite Dirac's equations (\ref{Dirac}) as
\begin{gather*}
\tilde{A}\partial_1\psi+\tilde{B}\partial_2\psi+\tilde{C}\psi-\lambda\psi=0,
 \end{gather*}
where
\begin{gather*}
\tilde{A}=
\begin{pmatrix}
 A_1 & A_2  \\
 -A_2 & -A_1  \\
\end{pmatrix},
\qquad
\tilde{B}=
\begin{pmatrix}
 B_1 & B_2  \\
 -B_2 & -B_1
\end{pmatrix},
 \qquad
 \tilde{C}=
\begin{pmatrix}
 C_1 &-C_2  \\
 C_2 & -C_1
\end{pmatrix},
\end{gather*}
and
\begin{gather*}
A_1=ie^1_1,\qquad A_2=-ike^1_2, \qquad B_1=ie^2_1, \qquad B_2=-ike^2_2,
\\
C_1=-\tfrac{i}{2}ke^\mu_2\Gamma^{12}_\mu, \qquad   C_2=-\tfrac{i}{2}e^\mu_1\Gamma^{12}_\mu.
\end{gather*}
Let $(x,y)$ be a coordinate system on the two-dimensional manifold and
\begin{gather*}	\psi=
\begin{pmatrix}
 \psi_1(x,y)  \\
 \psi_2(x,y)
\end{pmatrix}.
\end{gather*}
We now make the important assumption that the spinor $\psi$ is multiplicatively separable
\begin{gather*}\psi_i=a_i(x)b_i(y).
\end{gather*}
The Dirac equation then reads:
\begin{gather}
A_1\dot{a}_1b_1+A_2\dot{a}_2b_2+B_1a_1\dot{b}_1+B_2a_2\dot{b}_2+C_1a_1b_1-C_2a_2b_2-\lambda a_1b_1=0 ,\nonumber\\
-A_2\dot{a}_1b_1-A_1\dot{a}_2b_2-B_2a_1\dot{b}_1 -B_1a_2\dot{b}_2+C_2a_1b_1-C_1a_2b_2-\lambda a_2b_2=0. \label{eq:sys1}
\end{gather}
We now apply the general results of \cite{MR} to the Lorentzian case.

\begin{definition}
The Dirac equation \eqref{eq:sys1} and the operator ${\mathbb D}$ are separated in
$(x,y)$
 if there exists nonzero functions $R_i(x,y)$ such that \eqref{eq:sys1} can be rewritten as
\begin{gather}  R_1a_rb_s(E^x_1+E_1^y)=0,\qquad  R_2a_tb_u(E_2^x+E_2^y)=0 \label{eq:sys}
 \end{gather}
for suitable indices $r$, $s$, $t$, $u$ where $E_i^x(x,a_j,\dot{a}_j)$, $E_i^y(y,b_j,\dot{b}_j)$. Moreover, the equations
\begin{gather} \label{eq:cond}
 E_i^x=\mu_i=-E_i^y
\end{gather}
def\/ine the separation constants $\mu_i$.
\end{definition}

The above def\/inition refers to so-called ``naive'' separation of variables that is not always the most general.
In order to f\/ind symmetry operators, we adopt for our analysis some assumptions.

First of all, we assume a given coordinate system $(x,y)$ and impose a separation of \eqref{eq:sys1} according to the previous def\/inition. This can be done in three dif\/ferent ways
\begin{description}\itemsep=0pt
\item[Type I:] $a_1\neq a_2$ and $b_1\neq b_2$.
\item[Type II:] $a_1=a_2$ and $b_1\neq b_2$ (or viceversa).
\item[Type III:] $a_1=ca_2=a$ and $b_1=db_2=b$ ($c,d\in\mathbb C$).
\end{description}

Following the procedure laid out in \cite{MR}, we build eigenvalue-type operators $\mathbb{L}\psi=\mu\psi$ with eigenvalues $\mu(\mu_i)$ making use of the terms
$E_i^x$ and $E_i^y$ in \eqref{eq:cond} only. We require that the operators~$\mathbb{L}$ are independent of $\lambda$.
Furthermore, $\lambda\neq 0$.

Finally, we require the symmetry condition, that is $[\mathbb{L},\mathbb{D}]\psi=(\mathbb{LD}-\mathbb{DL})\psi=0$ for all $\psi$. A~operator~$\mathbb{L}$ which satisf\/ies the condition, is called a \emph{symmetry operator} since it maps solutions into solutions.

The symmetry operators are directly generated by the separated equations and having them enables one to immediately write down the same separated equations. In addition, the separation constants are associated with eigenvalues of symmetry operators.

The only relevant case is Type~I separation, since it is the only one associated with non-trivial second-order operators. We shall now make use of the naive separation assumption and of def\/inition~\eqref{eq:sys}. We would like to determine the indices $r$, $s$, $t$, $u$ and  $E_i^x(x,a_j,\dot{a}_j)$, $E_i^y(y,b_j,\dot{b}_j)$. Thus, we focus on what we can factorize from~\eqref{eq:sys1}.

By inspection, we notice that the only allowed factorizations are:
\begin{itemize}\itemsep=0pt
\item Factorize $a_1b_2$ in the f\/irst equation and $a_2b_1$ in the second equation.
\item Factorize $a_2b_1$ in the f\/irst equation and $a_1b_2$ in the second equation.
\end{itemize}
We consider the f\/irst factorization, which implies $A_1=B_2=0$. We divide into two parts both the equations: one part which is a function of~$x$ and the other a function of~$y$, in order to have~$E_1^x$,~$E_1^y$ and $E_2^x$, $E_2^y$.

Hence
\begin{gather*}
E_1^x=A_2R_1\frac{\dot{a_2}}{a_1}-C_2R_1\frac{a_2}{a_1}=\mu_1,\qquad
E_1^y=B_1R_1\frac{\dot{b_1}}{b_2}+C_1R_1\frac{b_1}{b_2}-\lambda R_1\frac{b_1}{b_2}=-\mu_1,
\end{gather*}
and
\begin{gather*}
E_2^x=A_2R_2\frac{\dot{a_1}}{a_2}-C_2R_2\frac{a_1}{a_2}=\mu_2,\qquad
E_2^y=B_1R_2\frac{\dot{b_2}}{b_1}+C_1R_2\frac{b_2}{b_1}+\lambda R_2\frac{b_2}{b_1}=-\mu_2,
\end{gather*}
where $\mu_1$ and $\mu_2$ are the separation constants.

The second-order operator ${\mathbb L}$ is therefore def\/ined by the following equations
\begin{gather*}
A_2R_1\dot{a_2}-C_2R_1a_2=\mu_1 a_1,\qquad
A_2R_1\dot{a_1}-C_2R_1a_1=\mu_2 a_2.
\end{gather*}
If we set $\mu_1\mu_2=\mu$, the operator is given by
\begin{gather*}
\mathbb{L}\psi:=\begin{pmatrix}
 A_2^2(x,y)R_1^2(y) & 0  \\
 0 & A_2^2(x,y)R_1^2(y)
 \end{pmatrix}\partial_x^2\psi\\
 \phantom{\mathbb{L}\psi:=}{} +
\begin{pmatrix}
 A_2R_1^2\partial_xA_2 -C_2A_2R_1^2-C_2R_1^2A_2 & 0  \\
 0 & A_2R_1^2\partial_xA_2  -C_2A_2R_1^2-C_2R_1^2A_2
\end{pmatrix}\partial_x\psi\\
  \phantom{\mathbb{L}\psi:=}{}
 +
\begin{pmatrix}
 \begin{matrix} C_2^2(x,y)R_1^2(y)\\ {}
 -A_2(x,y)R_1^2(y)\partial_xC_2(x,y)\end{matrix} & 0  \\
 0 &
  \begin{matrix}
 C_2^2(x,y)R_1^2(y)\\
 {} -A_2(x,y)R_1^2(y)\partial_xC_2(x,y) \end{matrix}
\end{pmatrix}\psi  =\mu\psi. 
\end{gather*}
We notice that the functions $A_2R_1$ and $C_2R_1$ are functions only of~$x$.

Finally, we look for the conditions that have to be applied in order to have the operator commuting with the Dirac operator~$D$: it results already that $[\mathbb{L},\mathbb{D}]=0$, so no other conditions are needed.  It follows from a detailed analysis of this case (D5 separation scheme in~\cite{MR}) that we obtain a Liouville metric with {\em one} ignorable coordinate.  This case will be discussed in detail in the next section. Another case is also possible which corresponds to the
D7 separation scheme in~\cite{MR}.  However, it may be shown that it is equivalent to previous one
modulo the sign of the Lorentz metric.

Separability for equations of Type~II and of Type~III  gives rise to f\/irst-order operators.

\section{Liouville metric}\label{section5}

In Section~\ref{section2} we concluded our analysis of symmetry operators with the condition \eqref{integra} on the second-order operator:
\begin{gather*}
\nabla^a g=-\tfrac{1}{4}\nabla_b(Re^{ab}).
\end{gather*}
Its analysis requires knowledge regarding which spin manifolds $M$ admit nontrivial valence two Killing tensors.

To further proceed with our analysis, we recall the following important result.

A two-dimensional Riemannian or Lorentzian space admits a non-trivial valence two Killing tensor if and only if it is a Liouville surface in which case there exists a system of coordinates $(u,v)$ with respect to which the metric $g$ and Killing tensor $K$ have the following forms
\begin{gather*}
g=(A(u)+B(v))\big(du^2+ \eta dv^2\big),
\\
K=\frac{B(v)}{A(u)+B(v)}\partial_u\otimes\partial_u-\eta \frac{A(u)}{A(u)+B(v)}\partial_v\otimes\partial_v=K^{ab}e_a\otimes e_b,
\end{gather*}
where $A$ and $B$ are arbitrary smooth functions. Furthermore, the frame component of $K$ are given by
\begin{gather*}
\big[K^{ab}\big]={\rm diag}\big(B(v),-\eta A(u)\big).
\end{gather*}
The spin frame corresponding to D5 separation discussed in the previous section is given by
\begin{gather*}
(e^{\mu}_a)=\begin{pmatrix}
 0 & \frac{1}{\sqrt{A(u)+B(v)}}  \\
 -\frac{1}{\sqrt{A(u)+B(v)}} & 0
\end{pmatrix},
\end{gather*}
where $A=0$ and $B(v)=1/R_1(v)^2$.

It follows from the previous section that the only Type~I separation, other than the equivalent one
discussed at the end of the previous section,
is associated with the nonsingular Dirac operator and associated symmetry operator of the form
\begin{gather*}
\mathbb{D}:=R_1(y)\left[   \begin{pmatrix}
 0 & k  \\
 -k & 0
 \end{pmatrix} \partial_x +
i   \begin{pmatrix}
 1 & 0  \\
 0 & -1
 \end{pmatrix} \partial_y\right]
 +\tfrac{i}{2}R'_1(y)   \begin{pmatrix}
 1 & 0  \\
 0 & -1
 \end{pmatrix},
 \qquad
 \mathbb{K}=
  \begin{pmatrix}
 \partial_x^2 & 0  \\
 0 & \partial_x^2
 \end{pmatrix}.
\end{gather*}
The operator $\mathbb{K}$ written above agrees with the second-order operator of Section~\ref{section3} computed for the Liouville metric
under the assumption $\alpha=\zeta=0$.  We observe that it is in fact the square of the f\/irst-order operator
\begin{gather*}
\mathbb{L}=   \begin{pmatrix}
 1 & 0  \\
 0 & 1
 \end{pmatrix} \partial_x,
\end{gather*}
corresponding to the
Killing vector associated to the ignorable coordinate $x$ of the Liouville metric
where, as we said before, $k=i$ for Euclidean and $k=1$ for Lorentzian signature.

The corresponding coordinates separate the geodesic Hamilton--Jacobi equation. If the mani\-fold is the Euclidean plane, the coordinates, up to a rescaling, coincide with polar or Cartesian coordinates.  In the Minkowski plane the coordinates correspond to pseudo-Euclidean or pseudo-polar coordinates.

\section{Separation in complex variables}\label{section6}

On real pseudo-Riemannian manifolds the Hamilton--Jacobi equation can be separated not only in standard separable coordinates but also in complex variables \cite{DR}. As in classical separation of variables theory, complex separable variables are determined by eigenvalues and eigenvectors of second-order Killing tensors. If the manifold is pseudo-Riemannian,  pairs of complex-conjugate eigenvectors and eigenvalues of second-order real Killing tensors can exist  in some part of the manifold, together with real ones. Where complex eigenvectors appear, it is impossible to determine real orthogonal separable coordinates, and the introduction of complex variables is necessary. The complex variables behave in all respects as complex coordinates, but they are not independent because of the conjugation relation. In the following their lack of independence will be irrelevant. Let us consider the 2-dimensional Minkowski  manifold with pseudo-Cartesian coordinates $(x,y)$. The geodesic Hamiltonian is given by
\begin{gather*}
H=\tfrac 12\big(p_x^2-p_y^2\big).
\end{gather*}
The space of valence two Killing tensors is 6-dimensional and there are ten dif\/ferent types of separable webs real in some part of the space~\cite{CDM}.
Another separable web, everywhere complex, is def\/ined by \cite{DR}
\begin{gather*}
z=x+iy, \qquad \bar z =x-iy,
\end{gather*}
and is determined by the eigenvectors of the Killing tensor whose non-null components in $(x,y)$ are
\begin{gather*}
K^{12}=K^{21}=1,
\end{gather*}
and the associated polynomial f\/irst integral is
\begin{gather*}
L=p_xp_y.
\end{gather*}
By def\/ining the canonical momenta as
\begin{gather*}
P=\tfrac 12(p_x-ip_y), \qquad \bar P=\tfrac 12(p_x+ip_y),
\end{gather*}
we have
\begin{gather*}
H=P^2+\bar P ^2, \qquad L=i\big(P^2-\bar P^2\big),
\end{gather*}
and a real complete separated  integral of the Hamilton--Jacobi equation can be determined~\cite{DR}. Because $(z,\bar z )$ are both ignorable variables, they should also separate the Dirac equation (indeed, in Minkowski space they are the only complex separable web with at least one ignorable variable). With respect to $(x,y)$ the Dirac operator can be written as
\begin{gather*}
{\mathbb D}=i\left[ \left( \begin{matrix} e^1_1 & -e^1_2 \vspace{1mm}\\ e^1_2& -e^1_1 \end{matrix} \right) \partial_x+\left( \begin{matrix} e^2_1 & -e^2_2 \vspace{1mm}\\ e^2_2& -e^2_1 \end{matrix} \right) \partial_y\right],
\end{gather*}
where $(e_a^\mu)$ are the components of the spin frame. Because the $(e_a^\mu)$ can be assumed to be constant, and since
\begin{gather*}
\partial_{z}=\tfrac 12(\partial_x-i\partial_y), \qquad \partial_{\bar z}=\tfrac 12(\partial_x+i\partial_y),
\end{gather*}
we have
\begin{gather*}
{\mathbb D}=i\left[ \left( \begin{matrix} e^1_1+ie_1^2 & -e^1_2-ie_2^2 \vspace{1mm}\\ e^1_2+ie^2_2& -e^1_1-ie^2_1 \end{matrix} \right) \partial_{z}+\left( \begin{matrix} e^1_1 -ie^2_1& -e_2^1+ie^2_2 \vspace{1mm}\\ e^1_2-ie^2_2& -e^1_1+ie^2_1 \end{matrix} \right) \partial_{\bar z}\right].
\end{gather*}
In order to write $\mathbb D$ in the form
\begin{gather*}
{\mathbb D}=\left[ \left( \begin{matrix} 0 & -1 \cr 1 & 0 \end{matrix} \right) \partial_{z}+i\left( \begin{matrix} 1 & 0 \cr 0 & -1 \end{matrix} \right) \partial_{\bar z }\right],
\end{gather*}
corresponding to Type~I separation in the Minkowski space, the components of $(e_a^\mu)$ in $(x,y)$ must be
\begin{gather*}
e^1_1=\tfrac i2, \qquad e_1^2=-\tfrac 12, \qquad e_2^1=\tfrac 12, \qquad e_2^2=-\tfrac i2 .
\end{gather*}
Therefore, by using $(e^\mu_a)$ and $(e_\mu^a)=(e^\mu_a)^{-1}$ for raising and lowering indices,
\begin{gather*}
(e_a)=\tfrac 12 (\partial_x+i\partial_y,\partial_x-i\partial_y).
\end{gather*}
It is remarkable that the spin frame base allowing the separation of variables in $(z,\bar z)$ is   essentially coincident with $(\partial _{\bar z}, \partial _{z})$.

Both $z$ and $\bar z$ are ignorable variables, therefore, both
\begin{gather*}
\left(\begin{matrix} 1 &0 \cr 0 & 1 \end{matrix}\right)\partial_{z}^2 \qquad \text{and} \qquad \left(\begin{matrix} 1 &0 \cr 0 & 1 \end{matrix}\right)\partial_{\bar z}^2,
\end{gather*}
can be used as dif\/ferential operators associated with separation of variables. The integration of the $\psi_i=a_ib_i$ is in all respects the same as for the separation in real coordinates.

\section{Appendix}\label{section7}

System~\eqref{cond} is obtained by using equations that characterize second- and third-order covariant derivatives. Such equations make use of Ricci's identities~(\ref{com1}) and~(\ref{com2}).

For second-order covariant derivatives the following equation hold
\begin{gather*}
\nabla_a\nabla_c\psi=
\tfrac{1}{2}
[\nabla_a,\nabla_c]\psi+\nabla_{ac}\psi=\nabla_{ac}\psi+\tfrac{R}{8}\epsilon_{ac}\gamma\psi. 
\end{gather*}
This last equation can be rewritten as
\begin{gather}
\nabla_c\nabla_a\psi=\nabla_a\nabla_c\psi+\tfrac{R}{4}\epsilon_{ca}\gamma\psi.\label{app2}
\end{gather}
Similar equations for third-order covariant derivative require more calculations and we will show only the main passages
\begin{gather*}
\nabla_{ab}\nabla_c\psi  =
\tfrac{1}{6}
(3\nabla_{a}\nabla_{b}\nabla_{c}+3\nabla_{b}\nabla_{a}\nabla{c})\psi\\
 \phantom{\nabla_{ab}\nabla_c\psi  }{}
 =\tfrac{1}{6} (\nabla_{a}\nabla_{b}\nabla_{c}+\nabla_{b}\nabla_{a}\nabla{c}+\nabla_{a}\nabla_{b}\nabla_{c}+2\nabla_{b}\nabla_{a}\nabla{c})\psi
\end{gather*}
Now by using \eqref{app2} and expanding:
\begin{gather*}
\nabla_{ab}\nabla_c\psi =\tfrac{1}{6} (\nabla_{a}\nabla_{b}\nabla_{c}+\nabla_{b}
\nabla_{a}\nabla_{c}+\nabla_{a}\nabla_{c}\nabla_{b}+\nabla_{b}\nabla_{c}\nabla_{a}+
\nabla_{a}\nabla_{c}\nabla_{b}+\nabla_{b}\nabla_{c}\nabla_{a})\psi\\
\phantom{\nabla_{ab}\nabla_c\psi =}+\tfrac{1}{12} \nabla_a R\epsilon_{bc}\gamma\psi+\tfrac{1}{12} \nabla_bR\epsilon_{ac}\gamma\psi
+\tfrac{1}{12} R\epsilon_{bc}\gamma\nabla_a\psi+\tfrac{R}{12} \epsilon_{ac}\gamma\nabla_b\psi.
\end{gather*}
Now the f\/irst part of the right hand side is just $\nabla_{abc}$. The second part can be rewritten to obtain
\begin{gather}
\nabla_{ab}\nabla_c\psi =\nabla_{abc}\psi+\tfrac{1}{12} (\nabla_a R\epsilon_{bc}+\nabla_b R\epsilon_{ac})\gamma\psi+\tfrac{R}{8} \epsilon_{ac}\gamma\nabla_b\psi\nonumber\\
\phantom{\nabla_{ab}\nabla_c\psi =}{}
+\tfrac{R}{8} \epsilon_{bc}\gamma\nabla_a+\tfrac{R}{12} (-\eta_{ac}\nabla_b-\eta_{bc}\nabla_a+2\eta_{ab}\nabla_c).\label{app3}
\end{gather}
Equations \eqref{app3} and \eqref{app2} together with Ricci's identities are all the tools needed to expand equation \eqref{sym1} in system \eqref{cond}.

\subsection*{Acknowledgements}

The authors wish to thank their reciprocal institutions, the Dipartimento di Matematica, Universit\`a di Torino and the Department of Applied Mathematics, University of Waterloo for hospitality during which parts of this paper were written.  The research was supported in part by a Discovery Grant
from the Natural Sciences and Engineering Research Council of Canada.

\pdfbookmark[1]{References}{ref}
\LastPageEnding

\end{document}